\documentclass[aps,showpacs,prb,twocolumn,floatfix]{revtex4}
\usepackage{amssymb}

\usepackage{graphicx}
\usepackage{amsmath}
\usepackage{srcltx}

\input{tcilatex.tex}
\input{tcilatex}

\begin{document}

\title{Comparison between two simple numerical models for the magnetoelectric
interaction in multiferroics}
\author{Cesar ~J. ~Calderon Filho and Gaston E. Barberis}

\address{Instituto de F\'{i}sica ''Gleb Wataghin'', UNICAMP,\\
13083-970, Campinas, S\~{a}o Paulo, Brazil.}

\begin{abstract}
We developed numerical calculations to simulate the magnetoelectric coupling
in multiferroic compounds, using the Monte Carlo technique. Two simple
models were used to simulate the compounds. In the first one, the magnetic
ions are represented by a spin 1/2 2D\ Ising lattice of ions, and the
electric lattice by classical moments, coupled one to one with the magnetic
moments. The coupling between both lattices allows to the leading lattice,
that is, the magnetic one, to change the orientation of the electrical
dipoles in one direction perpendicular to the magnetic dipoles. This
direction was chosen to accomplish the symmetry requirements of the
magnetoelectric effect. In the second case, the magnetic lattice is also a
2D Ising lattice, but the electric momenta are in a lattice that also
behaves as an Ising lattice, perpendicular to the magnetic moments. In this
case, the one-to-one coupling of the electric and magnetic momenta is
represented by a two-valued energy parameter, allowing the possibility of
independent transition temperatures for both lattices. Both models contain
three independent parameters. We studied the physical properties obtained
with both models, as functions of the ratio of the three parameters. The
results in both cases allowed us to compare changes in the physics of the
models, and with the physics of compounds measured experimentally.
\end{abstract}

\date{\today }
\maketitle

\section{Introduction.}

Recently, there has been a revived interest in the research of multiferroics
due to several new discoveries, and the possibility to use them
technologically.$^{1}$ The electric and magnetic transitions are not
necessarily correlated, but when it occurs - and the so called \textit{%
magnetoelectric effect }appears - the materials suggest possible use as
memories, etc. Besides the technical applications, several families of those
compounds present very rich physics. Just as an example, one of this
compounds, LiNiPO$_{4}$, shows a phase transition where the electric lattice
is not only first order, but in a very short range of temperature several
incommensurable transitions appear.$^{2}$ Even in front of those interesting
phenomena, we found very few theoretical papers on this subject, and most of
them using very elaborated theoretical methods.$^{3}$

With the motivation presented above, we developed two simple numerical
models, based on the Monte Carlo method, and the Metropolis minimization of
total energy.

We present here two simple and understandable models for the
magnetoelectricity. The electromagnetic Hamiltonian is solved for very
simple cases, and the solutions present similitude with the reported
experiments. The first model study the phase transitions at the same
temperature, independently of the temperature value. The second model, a
little more elaborated, allows the ferroelectric and ferromagnetic
transitions to occur at different temperatures, and we studied the behavior
of the model as function of this temperature difference. We compare the
results between the models and with experimental cases.

\section{The models.}

The physics behind the magnetoelectric effect consists, as seen in a bird's
eye view, \ in the creation or orientation of electric dipoles by the
magnetic moments or vice-versa. In the first case, the magnetic dipoles,
which are permanent, when change their orientation, modify the lattice in
such a way that the negative electric charges displace relatively to the
positive. This is accomplished via spin-orbit coupling of the spins,
changing the total energy as the orbit lattice Hamiltonian explains.
Simplifying the model, the spin-orbit-lattice energy is calculated as a
spin-lattice Hamiltonian. We simplified the calculation even more,
representing the magnetic lattice as a 2D Ising lattice in all the cases.
This corresponds with many real compounds, as the olivines mentioned above,
where the structure of the real lattice presents separated planes of
magnetic ions.$^{4-6}$

We assume that the magnetoelectric system is a set of magnetic dipoles,
coupled via the exchange interaction, in a lattice with a distribution of
electric charges, susceptible to change when the magnetic dipoles change
their orientations. The change in orientation of the magnetic dipoles modify
their environment, via spin-orbit interaction, creating local strains, and
creating or orienting a set of electric \ dipoles in the lattice. We assume
that our crystal suffers the strain in such a way that electric dipoles are
oriented to a particular direction when the magnetic dipoles relax.

The model Hamiltonian used in the models follows:
\begin{equation}
H=H_{M}+H_{E}+H_{ME}
\end{equation}
where $H_{M}$ is the magnetic energy, $H_{E}$ the electric energy and $%
H_{ME} $ the magnetoelectric coupling.

\subsection{\protect\bigskip The first model}

The first approach for a solution of eq.(1) is obtained replacing the first
term in the sum of the right side by a square sublattice of Ising magnetic
moments, and the electric moments in the second and third terms by random
oriented classical electrical dipoles, located in a separated square
sublattice. The Ising spins are coupled to their nearest neighbors only, and
with periodic boundary conditions. The interaction Hamiltonian allows only
nearest neighbors magnetoelectric interaction. Symmetry requires that the
magnetic point group of the magnetic moment is one of the 58 Shubnikov
groups that allow magnetoelectricity.$^{7}$ This forces our magnetic moments
to have only one electric dipole as a nearest neighbor. The electromagnetic
coupling is divided in two parts: the \textit{local }interaction between the
spin and the electric dipole, and the \textit{lattice }total electromagnetic
energy, that takes into account the interaction between the electric dipoles
and their electric neighbors. As most of the electric parameters are
measured perpendicularly to the magnetization,$^{2,8}$ we chose the $%
\widehat{z}$ direction for the magnetic moments, and the $\widehat{x}$ axis
for the electric dipoles.

The numerical solution of the problem was done using the importance sample
Monte Carlo method, looking for the minimum in energy for our system.

Thus,

\begin{equation}
H=-J\sum_{<i,j>}\sigma _{i}\sigma _{j}-h\sum_{i}\sigma _{i}-\beta
\sum_{\{i,j\}}P_{ix}P_{jx}+\gamma \sum_{i}P_{ix}
\end{equation}

\bigskip

is the approximated Hamiltonian, where $J$ is the exchange coupling of the \
Ising magnetic spins $\sigma $, and $P$ the electric dipoles. The symbol $%
<i,j>$ indicates sums over the nearest neighbors only. The first and second
terms constitute the magnetic energy, where we included the possibility of
an applied or external magnetic field $h$. The third term represent the
electric energy, proportional to the orientation of the neighboring electric
momenta. As the system is to be ferroelectric, and the direction of the
polarized dipoles the $\hat{x}$ axis of the crystal, we considered only the
energy coupling in that direction, which is represented by $\{i,j\}$,
indicating sum over the two first neighbors located in the $\hat{x}$ axis.

The interaction term, which represents a spin - lattice Hamiltonian, was
separated into two parts. One of them is the local interaction between the
spin and the $\widehat{x}$ projection of the electric dipole, which makes
that every transition of the spin changes simultaneously the dipole; the
second, represented the last term in eq. (2) is the contribution of the
lattice as a whole to the total energy. As the local interaction is the same
for every pair spin-dipole, we did not include it in the Hamiltonian.
However, the meaning of this local part of the energy is important, as we
will see below.

\subsubsection{\protect\bigskip Ferromagnetic case.}

Our first calculation was performed in a 100$\times $100 2D lattice of Ising
ferromagnetic spins coupled to 100$\times $100 electric dipoles, located in
another square lattice, parallel to the magnetic one, and slightly shifted
from it. The electric dipoles were oriented at random, together with
magnetic lattice, to begin with infinite temperature. The temperature was
then fixed to a value, and a Monte Carlo program, where the transitions are
allowed following the Metropolis technique,$^{9}$ is iterated the time
necessary to obtain thermal equilibrium of the system. Then, the results
are used as \ the initial condition for the following temperature. The
calculation was performed reducing the temperature in each step.

The complete calculation was performed after a study of convergence in our
model. As first step, we unconsidered the electric and interaction energies
when we looked for the minimum. This means that the model is just a 2D Ising
system, moving electric dipoles together, and as expected, the magnetization
follows the Ising model. The exact calculation published by Onsager allows a
very good comparison, and the electric dipoles also feel a transition at the
same temperature. This calculation decided the size of the set of moments,
which we selected as 100x100 on this basis.

The following step was to study the convergence when the parameters $\beta $
and $\gamma $ in the model are different from zero. It is well known that
the Ising model converges slowly near by the transition temperature, due to
fluctuations, and the equal value for the energy when the system is oriented
in any of the both possible directions. This is easily solved with the
addition of the small external field $h$; however, we observed that for
particular values of $\gamma $, the convergency is the slowest. This can be
explained by the fact that $\gamma $ appears in the Hamiltonian as an extra
external field, in some manner. We can use the local coupling of the
spin-dipole pair to write
\begin{equation}
P_{ix}=P_{ix}|\sigma _{i}|=\sigma _{i}|P_{ix}|
\end{equation}
for the modulus of $\sigma _{i}$ is always the unity; that can be used to
write the second and last terms in the Hamiltonian as

\bigskip

\begin{equation}
\gamma \sum_{i}P_{ix}-h\sum \sigma _{i}=-\sum (h-\gamma |P_{ix}|)\sigma _{i}
\end{equation}
which shows that the value of $\gamma |P_{ix}|$ appears as an extra magnetic
field. The mean value of $|P_{ix}|$ annulate the external field when $\gamma
/J\approx 0.02$ for an applied field $h/J=0.01$, and the convergency is the
slowest for this value of $\gamma $. Taking this into account, we found that
it was necessary 5000 iterations per spin to obtain thermal equilibrium and
other 5000 to get the mean values of the energy and magnetization; this
numbers were used in every case.

Eq. (4) may also be used to analyze the meaning of the $\beta $ parameter.
The first and third terms in eq.(2) can be written
\begin{eqnarray}
H_{1} &=&-J\sum_{<i,j>}\sigma _{i}\sigma _{j}-\beta
\sum_{\{i,j\}}P_{ix}P_{jx} \\
&=&-J\sum_{<i,j>}\sigma _{i}\sigma _{j}-\beta \sum_{\{i,j\}}|P_{ix}|\sigma
_{i}|P_{jx}|\sigma _{j}  \notag \\
&=&-\sum_{\{i,j\}}(J+\beta |P_{ix}||P_{jx}|)\sigma _{i}\sigma
_{j}-J\sum_{<i,j>\neq\{i,j\}}\sigma _{i}\sigma _{j}  \notag
\end{eqnarray}

So it can be seen that the $\beta $ parameter makes the exchange
anisotropic, modifiyng its value in the $\widehat{x}$ direction, leaving the
coupling in the $\widehat{y}$ direction unaltered.

We performed the calculation as function of the temperature for different
values of the $\beta $ parameter for $\gamma =0$; then repeated the
calculation for $\beta =0$ to study the dependence of the results from $%
\gamma $, and finally we made a complete study of the form and transition
temperature of the system as a function of both parameters. As $J$
determines the transition temperature for the Ising model, we used it as
unit of energy for the whole system.

To make the results clear, we made calculations as function of the
temperature for different values of $\beta $, when $\gamma $ is zero -
meaning that the electric interaction is bigger than the magnetoelectric
one. After that, we calculated the minimum as function of $\gamma $, when $%
\beta $ is zero. The complete calculation, with both parameters different
from zero gives a clear vision of the total behavior of our model. The
results for the ferromagnetic case are shown in Figs. 1, 2, and 3. The first
and second figures show the change in the shape of the transition caused by
the $\beta $ and $\gamma $ parameters. The effect of the$\ \beta $ parameter
is principally to shift the transition, as seen in Fig. 2. The $\gamma $
value can be positive and negative, and the effect is to broaden the
transition, and invert the magnetization-polarization when the sign of it is
changed. Fig. 3 shows the complete dependence of the transition temperature
depending of both parameters.
\begin{figure}[tbp]
\includegraphics[scale=0.30]{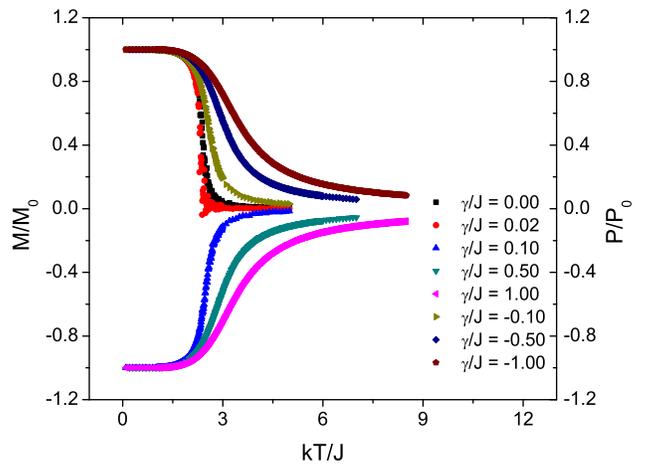}
\caption{(color online) Normalized magnetization (and electric polarization)
as function of the temperature when $\protect\beta =0$ in the first model.}
\label{Figure 1}
\end{figure}
\begin{figure}[tbp]
\includegraphics[scale=0.30]{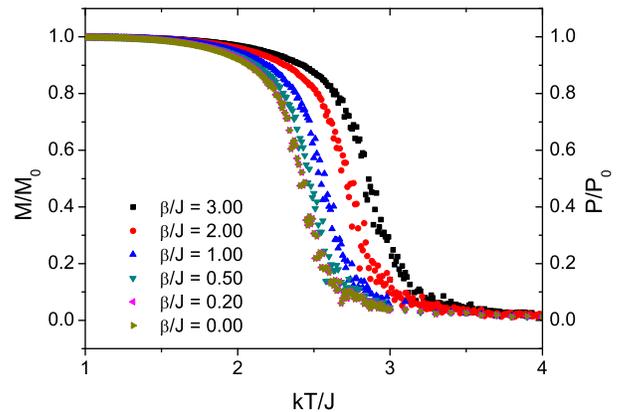}
\caption{(color online) The same temperature dependencies as in Fig. 1, when
$\protect\gamma =0,$ in the first model.}
\label{Figure 2}
\end{figure}

\begin{figure}[tbp]
\includegraphics[scale=0.30]{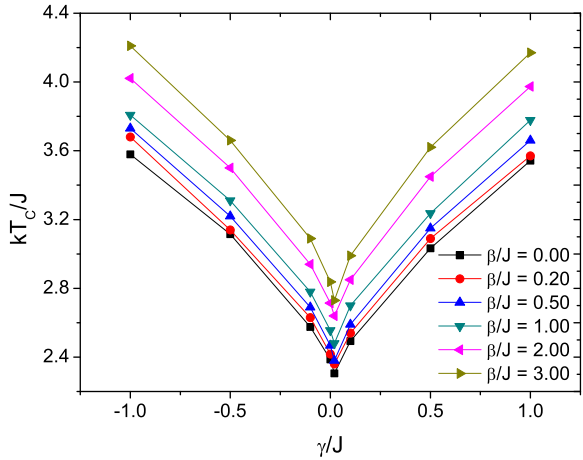}
\caption{(color online) The general results for the first model. The
transition temperature is the same for both the electric and the magnetic
transitions (see text).}
\label{Figure 3}
\end{figure}

Figs. 1, 2 and 3 show the results for the ferromagnetic case. Fig. 1 shows
the effect of the $\gamma $ parameter: as the transition is determined by $J$%
, the effect of $\gamma $ is to broaden the transition. Fig. 2 shows the
effect of $\beta $, which is to shift the transition without great
modifications in the shape of it. Fig. 3 resumes the complete model, showing
the effect of both parameters together. The induced electric polarization
appears at the same temperature as the magnetic transition in all the cases,
as it is defined by the model. As the model only allows, both $P/P_{0}$ and $%
M/M_{0\text{ }}$curves coincide.

\subsubsection{Antiferromagnetic case}

The antiferromagnetic case was treated similarly. It requires a negative
value of $J$, but several changes in the other terms of the Hamiltonian are
necessary. The magnetization of both sublattices will be coupled to the
electric lattice opposed, in order to obtain the required ferroelectricity.
The magnetic field is set to zero, because only can be directed parallel to
one of the magnetic sublattices. The convergence is slowest for zero field,
which we used to obtain the values for convergence. Our results in this case
are very similar to those above, and we lack here of space to show them
completely. Fig. 4 presents the transition temperature dependence for this
case, which can be compared with the ferromagnet. The complete results will
be publish elsewhere, together with \ a more sophisticated model for the
spin-lattice coupling.
\begin{figure}[tbp]
\includegraphics[scale=0.30]{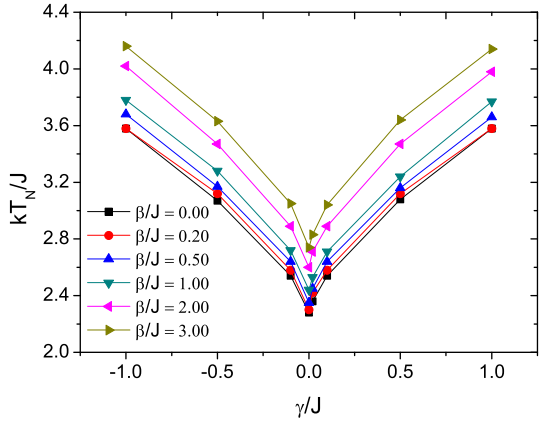}
\caption{(color online) The general results for the first model in the
antiferromagnetic case. The transition temperature is the same for both the
electric and the magnetic transitions (see text).}
\label{Figure 4}
\end{figure}
\bigskip
\begin{figure}[tbp]
\includegraphics[scale=0.3]{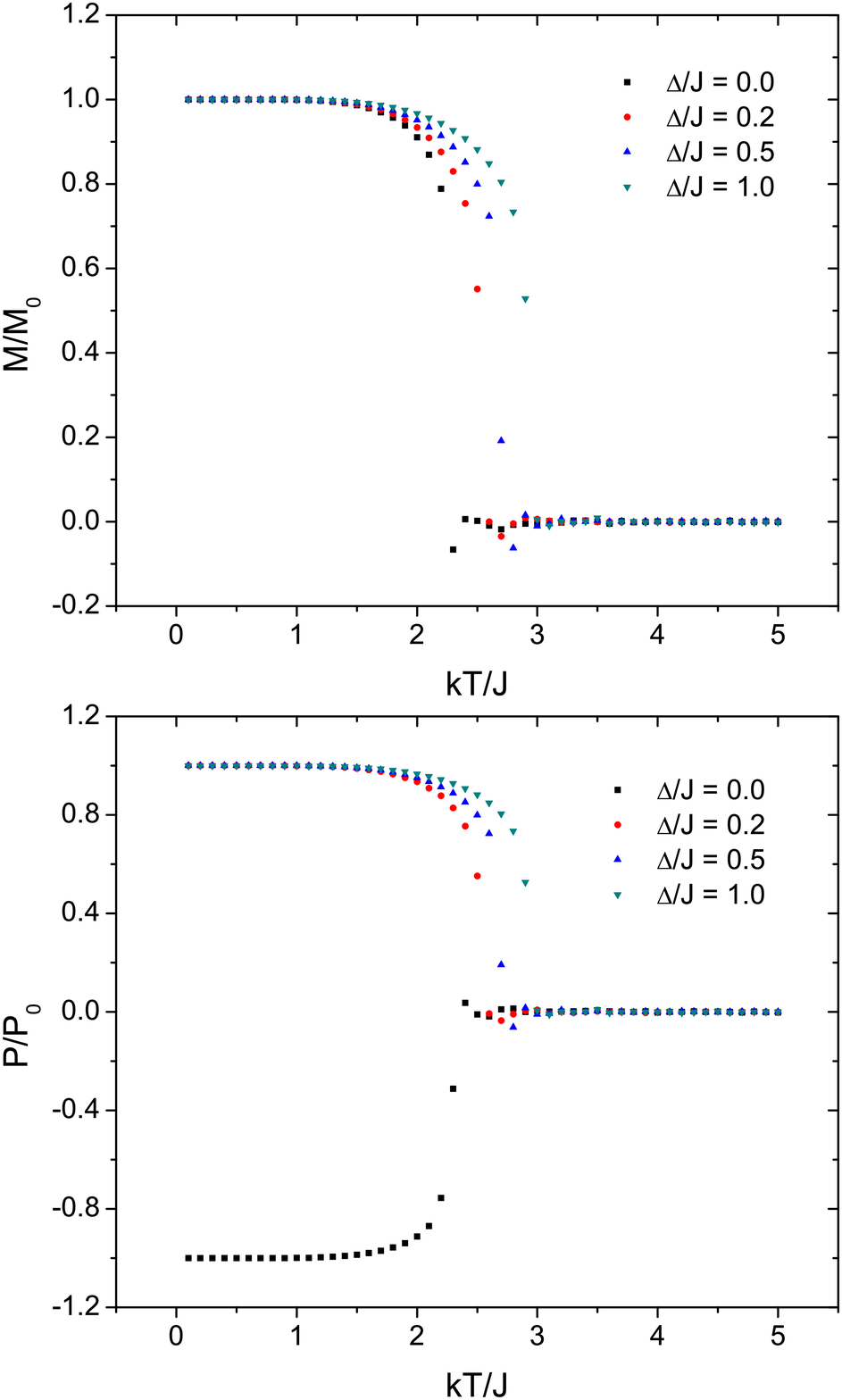}
\caption{(color online) Normalized magnetization and electric polarization
as functions of T for $\protect\beta /J=1$ in the second model.}
\label{Figure 5}
\end{figure}
\bigskip
\begin{figure}[tbp]
\includegraphics[scale=0.3]{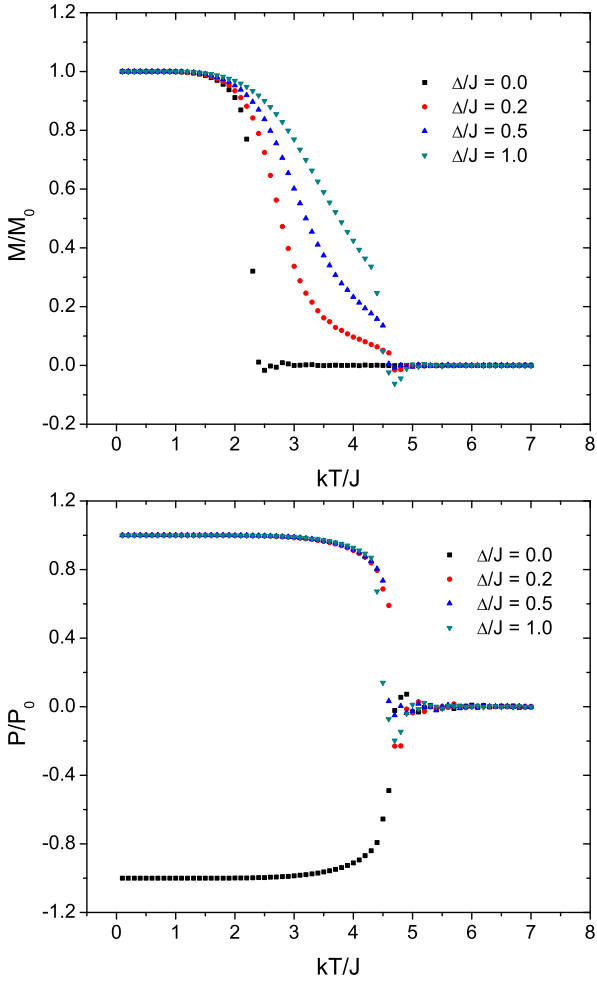}
\caption{(color online) Calculated magnetization and polarization per
spin/dipole, for the second model when $\protect\beta /J=2$ as functions of
T. It can be seen that M/M$_{0}$ is strongly distorted, while P/P$_{0}$ is
not.}
\label{Figure 6}
\end{figure}
\bigskip
\begin{figure}[tbp]
\includegraphics[scale=0.3]{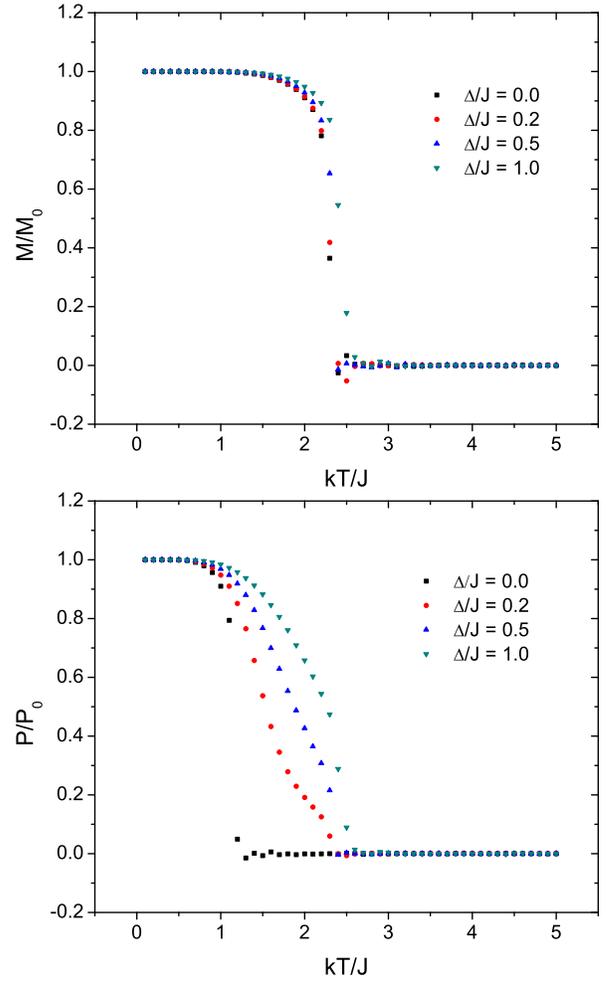}
\caption{(color online) Results for $\protect\beta /J=0.5$ in the second
model. In this case, the magnetic function is not distorted, but the
polarization is.}
\label{Figure 7}
\end{figure}
\bigskip
\begin{figure}[tbp]
\includegraphics[scale=0.3]{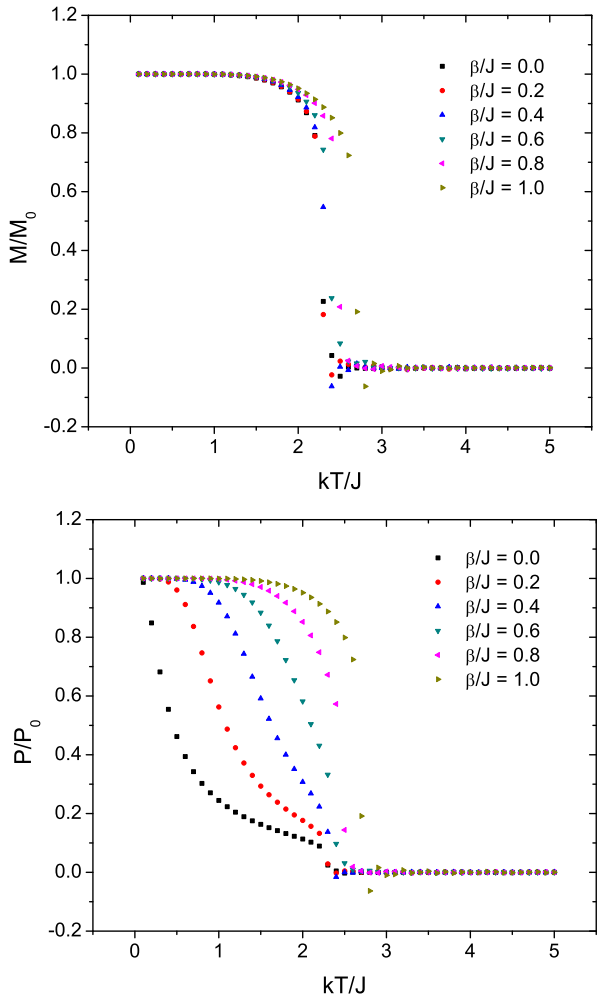}
\caption{(color online) Results of the second model when $\Delta /J=0.5$ and
changing the $\protect\beta $ parameter for $\protect\beta /J\leq 1$ . The
shape of the polarization curves are distorted and shifts as the electric
transition temperature is increased.}
\label{Figure 8}
\end{figure}
\bigskip
\begin{figure}[tbp]
\includegraphics[scale=0.3]{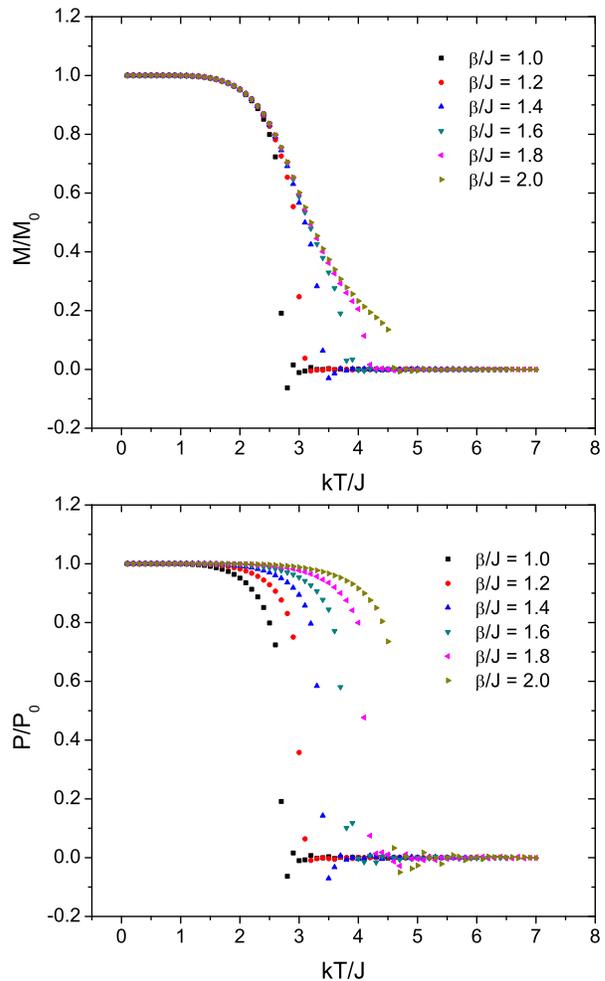}
\caption{(color online) Calculated polarization and magnetization for $%
\Delta /J=0.5,$ for values of the $\protect\beta $ parameter when $\protect%
\beta /J\geqslant 1.$}
\label{Figure 9}
\end{figure}
\bigskip
\begin{figure}[tbp]
\includegraphics[scale=0.3]{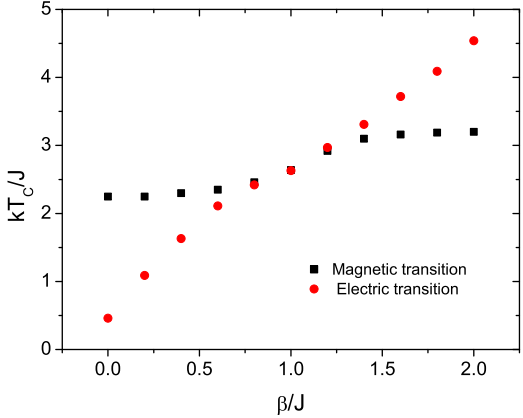}
\caption{(color online) Electric and magnetic transition temperatures, for $%
\Delta /J=0.5$ as depending of the $\protect\beta $ parameter. }
\label{Figure 10}
\end{figure}
\bigskip
\begin{figure}[tbp]
\includegraphics[scale=0.3]{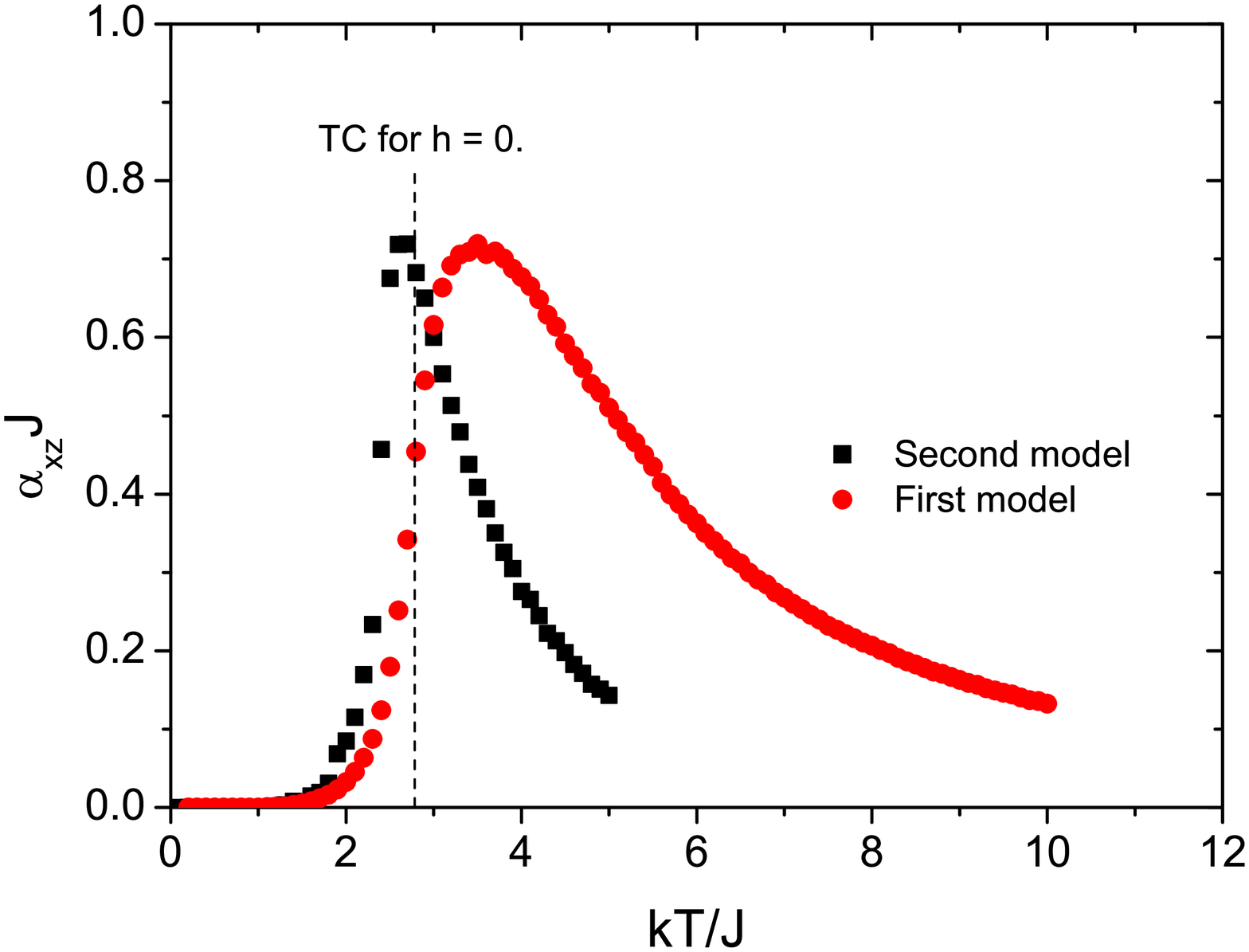}
\caption{(color online) The magnetoelectric coefficient as calculated for
our models. As expected, they tend to zero at T=0, as the magnetizations and
polarizations saturate. }
\label{Figure 11}
\end{figure}
\bigskip

\subsection{The second model}

As seen above, the first model does not contain the capacity to allow
different temperatures for the electric and the magnetic transitions, giving
us only the changes generated in the shape of the transitions by the
magnetoelectric coupling. We decided to develop a second model, where the
transition temperatures are independent. To maintain the simplicity of the
model, and the number of independent parameters reduced to three, we
included an energy $\Delta =\varepsilon _{2}-\varepsilon _{1}$ for the pair
of magnetic-electric moments. When the spin is up, and the electric dipole
points the left, or when the spin points down, and the electric dipole to
the right, they will have an energy $\Delta $ higher than in the other two
cases. If we make $\Delta \rightarrow \infty $ we recover the first model
again, for the system will be in the lower state all time.

We changed some other things in the new model. Instead of the classical
electric dipoles oriented at random at $T\rightarrow \infty $, we
substituted the electric lattice by an Ising lattice, oriented in the $%
\widehat{x}$ direction, that is, as formerly, the electric dipoles form the
ferroelectric part of the lattice when they are oriented in the positive $%
\widehat{x}$ direction. We excluded the local interaction between the pairs,
considering that the electric interaction is only in the $\widehat{x}$
direction, let's say, the tails of the interaction after first neighbors
cutted off. The result is that we substitute the magnetoelectric interaction
for this two-level system. Hence, the complete Hamiltonian is, in this case
\begin{equation}
H=-J\sum_{<i,j>}\sigma _{i}\sigma _{j}-h\sum_{i}\sigma _{j}-\beta
\sum_{<i,j>}P_{ix}P_{jx}+\sum_{i}\varepsilon _{i}
\end{equation}
where the symbols are the same as before, and the $\varepsilon _{i\text{ }}$%
the energy of the pair magnetic-electric momenta as defined above. We used
again the exchange coupling parameter $J$ as energy unit; as can be seen ,
this model has two independent transition temperatures for the magnetic and
electric lattices, as $J$ and $\beta $ are independent in this model.

To use the Monte Carlo method again, we need to preserve the mathematical
requirements for it, then our model needs to behave as a Markovian one. To
accomplish this requirement, the minimum of energy is calculated through the
following steps:

a - As in the first model, the initial state of the system is at $%
T\rightarrow \infty $. Then the value of the temperature is inserted in the
calculation.

b - One number $A$ is then chosen at random between zero and one. If this
number is zero, we invert the corresponding spin; instead, for $A=1$, the
inversion is performed on the electric dipole.

c - One of the pair of moments is chosen at random. If $A=0$, we calculate
the energy difference if the spin is inverted. From eq. (6), this energy
difference will be
\begin{equation}
\Delta E_{1}=2\sigma _{i}(JS_{S}+h)+\sigma _{i}P_{ix}\Delta
\end{equation}
where $\sigma _{i}$ is the chosen spin, $S_{S}$ the sum of the first \ spins
which are nearest neighbors to it, and $P_{ix}$ the component of the
electric dipole of the pair. If $\Delta E_{1}$ is negative, the spin is
inverted. If $\Delta E_{1}$ is positive, we use the Metropolis comparison:
if a new random number $0\leq r$ $\leq 1$, when compared with the energy
population is less than that (that is $r\leq \exp \left( -\Delta
E_{1}/k_{B}T\right) $) the spin is inverted too. If not, the spin is left
unaltered.

d - If $A=1$ the electric dipole is inverted; the energy difference in this
case is
\begin{equation}
\Delta E_{2}=2P_{ix}\beta S_{D}+\sigma _{i}P_{ix}\Delta
\end{equation}
Here the value of $S_{D}$ is the sum over electric dipoles which are nearest
neighbors to the one chosen. The other symbols have the meaning above.
Again, if $\Delta E_{2}$ is negative, the dipole is inverted, and if
positive, the random number $r$ is calculated and used to compare with the
populations in order to decide the inversion of the dipole.

e - The procedure from b) to d) is repeated as many times as necessary to
obtain convergence to thermal equilibrium as defined for the first model.
Then the temperature is reduced further and the calculation repeated to
equilibrium.

The model respects the symmetry requirements for magnetoelectricity. The
system does not contains temporal nor spacial inversion, so this exigency is
accomplished too.

\subsubsection{Convergence.}

Differently from \ the first case, this second one does not contain a strong
local coupling for the pair, and even the local change is not always decided
by the spin system. It was necessary, then, to realize an independent study
for every pair of the parameters, $\Delta $ and $\beta $. We do not believe
that the insertion of the whole convergence study could be interesting for
the reader, so we just mention that the number of Monte Carlo steps per spin
for convergency varies from 5 to 50 thousand steps, the small value for $%
\Delta /J=0,\beta /J=1$ and the maximum for $\Delta /J=1,\beta /J=2.$

\subsubsection{Results}

We performed complete calculations for several cases, described below

1 - Thermal dependence of polarization and magnetization when $\beta /J = 1$
(that is, the transitions occur at the same temperature) and $\Delta $
varies.

2- The same study, for $\beta /J>1.$

3 - The same again, when $\beta /J<1.$

4 - The thermal dependence, as function of $\beta ,$ for $\Delta /J=0.5.$

\bigskip Now we will describe the results for every case from 1 to 4.

1 - As the electric dipoles are not strongly coupled to his magnetic
neighbor, and they can assume both orientations, the shapes of the
transitions are different, as can be seen in Fig. (5). As there is not an
applied electric field, the relative orientation of the polarization and the
magnetization could be at random, as can be seen in the figure. The
transition temperature increases with the value of $\Delta ,$ meaning that
the coupling helps to mantain the system ordered.

2 - Fig. (6) shows the results for $\beta /J=2.$ As can be seen, the
transitions differ strongly in shape, and the curve of magnetization is
shifted to accompany the polarization transition, as $\Delta $ increases.
This coincides with the idea that the second model limits with the first,
when $\Delta \rightarrow \infty $. Anyway, we were not able to calculate
this case for high values of $\Delta ,$ for the time of convergence
increases too much.

3 - The case where the electric transition occurs at lower temperature than
the magnetic one is presented in Fig. (7). Here, the magnetic transition is
not deformed, and the electric one is distorted, with the tendency to
accompany the magnetization for $\Delta \rightarrow \infty .$ Again, we were
limited to values of $\Delta $ allowed by the time of convergency of the
calculation.

4 - The calculation was performed for $\Delta /J=0.5$ for different values
of the transition temperatures ($J\lessgtr \beta $). The results are shown
graphically in Figs. (8) and (9). We separated the results when the electric
transition is at lower temperature than the magnetic (Fig. (8)), and the
opposite, when $\beta \geq J$ (Fig. (9)). It can be observed in both cases that
the transition occuring at higher temperature remains almost undistorted,
and the one whose transition temperature is smaller, distorced and shifts.
Fig. (10) shows the transition temperatures, eletric and magnetic, when $%
\Delta /J=0.5$ as functions of $\beta .$ Of course, both curves equal when $%
\beta /J=1.$ The magnetic transition tends to saturate for small or great
values of $\beta ,$ while the electric transitions behave almost
linearly.\\

\section{Conclusions and perspectives.}

Magnetoelectricity and magnetoferroics are studied experimentally using the
phenomenological free energy obtained just from symmetry and the (possible)
interaction \ between magnetic and electric fields, as follows$^{1}$:
\begin{eqnarray*}
F(\overrightarrow{E},\overrightarrow{H})
&=&F_{0}-P_{i}^{S}E_{i}-M_{i}^{S}H_{i}- \\
&&-\frac{1}{2}\epsilon _{0}\epsilon _{ij}E_{i}E_{j}-\frac{1}{2}\mu _{0}\mu
_{ij}H_{i}H_{j}-\alpha _{ij}E_{i}H_{j}-...
\end{eqnarray*}
and we have:
\begin{eqnarray*}
P_{i} &=&-\frac{\partial F}{\partial E_{i}}=P_{i}^{S}+\epsilon _{0}\epsilon
_{ij}E_{j}+\alpha _{ij}H_{j}+... \\
M_{i} &=&-\frac{\partial F}{\partial H_{i}}=M_{i}^{S}+\mu _{0}\mu
_{ij}H_{j}+\alpha _{ji}E_{j}-...
\end{eqnarray*}
where it can be seen that the experimental measurement of the
magnetoelectric tensor, $\alpha _{ij},$ is performed looking for the
difference of the observed magnetization (polarization) with and without an
external magnetic (electric) field. To calculate that parameter, we followed
the same procedure, calculating for every temperature the magnetic
polarization with and without an applied field, that is
\begin{equation*}
\alpha _{ij}(T)\backsimeq \frac{P_{i}-P_{i}^{S}}{H_{j}}
\end{equation*}
as the experiments are made.$^{3}$

Fig. (11) presents the results for both models, when the transition
temperatures are the same ($J=\beta $) in the second model. The
magnetoelectric coefficients \ go to zero at T = 0, which is expected in a
system that saturates magnetically and electrically too. Our models do not
include the possibility to change the energy saturated at 0 K, but the
experiments (see ref. 3) show a remanescent value of the parameter. Our
model does not include any effect in other than the $\widehat{x}$ axis,
thus, we only obtained the $\alpha _{xz}$ magnetoelectric coefficient within
both models.

The transition in ref. 3 is first order, that is, the coefficient does not
exist for temperatures above the magnetic transition, contrary to our
models, where both transitions are second order, and as that, the curves in
Fig. (11) extend to high temperatures.

Resuming, our calculation arrives to many similitudes and differences with
experiment. We believe that this can be the simpler way to simulate real
systems, and developing more elaborated spin - lattice terms in the
Hamiltonians will help to interprete the experimental results in an easy way.

\ 


\end{document}